# Appreciable magnetic moment and energy density in single step normal route synthesized MnBi

Nithya R. Christopher[1,2], Nidhi Singh[2], Shiva Kumar Singh[2], Bhasker Gahtori[2], S.K. Mishra[1], A. Dhar[2] and V.P.S Awana[2*]

[1]Research and Technology Development Centre, Sharda University, Greater Noida, U.P. - 201306, India

[2]National Physical Laboratory (CSIR), New Delhi-110012, India

## Abstract

We study the structural and magnetic properties of the MnBi inter-metallic compound. The LTP (Low Temperature Phase) MnBi compound is successfully synthesized in single step by vacuum encapsulation technique and rapid quenching from phase formation temperature. The phase purity and the magnetic moments of MnBi are highly dependent on heat treating schedule. The best phase purity and the magnetic moment are found for a sample heat treated at 310$^o$C for 48h. Rietveld fitted X-ray diffraction (*XRD*) patterns revealed that the studied MnBi compound is crystallized in hexagonal *P63/mmc* space group with minute presence of unreacted Bi and Mn phases. The scanning electron microscopy (*SEM*) study is carried out to visualize the grains morphology and phase identification. The bulk MnBi powder showed appreciable magnetic moment of (~ 62emu/g at 6Tesla) and maximum energy product (BH$_{max}$ of 4.01MGOe at 6Tesla). The magnetic properties of synthesized MnBi show that it could be a potential candidate for rare earth free permanent magnets.

Key words: Permanent magnets, Structural refinement, Magnetization and Magnetic energy.

*Corresponding author's email: awana@mail.nplindia.ernet.in,

Fax No. +91-11-45609310: Phone no. +91-11-45609357

*Web page*- *www.freewebs.com/vpsawana/*

## 1. Introduction:

The scientific community is focusing on rare earth free permanent magnetic materials as they can be potential substitutes for rare earth based magnets [1-4]. This is due to the fragile supply chain, responsible for providing adequate amounts of rare earth materials, making rare earth based magnets rare and expensive [5]. In this regard Mn-based inter-metallic materials such as MnBi [6] and MnAl [7] are attracting a lot of attention of materials scientist around the globe. These compounds also show good magneto-optical properties and hence could be used as permanent magnetic materials for new thermal writing media [8-9]. Pure MnBi crystallizes in the NiAs-type hexagonal crystal structure at room temperature having $c$ axis as their easy axis of magnetization [6, 9]. According to the synthesis temperatures, MnBi binary system exhibits three magnetic phases. The MnBi synthesised below 355$^o$C, shows ferromagnetic behaviour. This phase is known as ferromagnetic low temperature phase (LTP-FM) [10]. The phase diagram of MnBi is interesting that above 355$^o$C, it undergoes first order structural (NiAs-type to distorted $Ni_2In$-type hexagonal) and magnetic transition to a paramagnetic high temperature phase (HT-PM) accompanied by a phase decomposition of MnBi into $Mn_{1.08}$Bi [8, 11]. When HTP-MnBi is rapid quenched, a ferromagnetic quenched high temperature phase (QHTP) with the curie temperature of 167$^o$C is obtained [6]. The MnBi binary system possesses favourable magnetic and magneto-optical properties [9].

The inter-metallic compound MnBi in its LTP phase exhibits attractive hard magnetic properties, i.e., high uni-axial magneto-crystalline anisotropy ($K$ ≈$10^7$ergs/cm$^3$) and an unusually high positive temperature coefficient of coercivity [9, 12-14]. It is remarkable that the coercivity of the LTP-FM increases with temperature, exhibiting a maximum value of 1.9 Tesla at 277°C [9]. In view of the above mentioned interesting properties, the LTP-FM phase of MnBi has very good potential as a permanent magnetic material. The crystal structure, orientation and magnetic properties of MnBi phases depend upon the method of preparation and the heat treatment. Xu.et.al reported simultaneous phase formation of various (LTP-FM/HT-PM/QHTP) phases of MnBi [15]. Single-phase MnBi is difficult to obtain using conventional methods, such as arc melting or rapid solidification techniques [16-18]. This is due to the immediate formation of Mn and Bi precipitates, within the required MnBi matrix. Preparation of single phase MnBi in large amounts remains difficult due to segregation of Mn above peritectic temperature i.e., 446$^o$C [17, 18]. Lot of research work has been done to synthesize single phase MnBi [16, 19-20]. Yang et al. obtained 90 wt. % LTP-FM MnBi magnets by magnetic separation and field aligned resin binding [21]. Guo et.al prepared the MnBi magnet by rapid quenching, followed by a thermal treatment and obtained more than 95 wt. % LTP-FM phase [14]. Adams *et.al* synthesized MnBi magnet by a hot pressing technique and obtained the LTP-FM phase with maximum energy product [$(BH)_{max}$] of 4.3MGOe and relative density of 90% [22]. Recently, Zhang *et.al* prepared 93%-dense MnBi permanent magnets by spark plasma sintering (SPS) of mechanically milled powders and obtained remanence ($M_r$) of 30emu/g [23]. Xu.et.al synthesized melt spun MnBi ribbons, but this method often produces non-crystalline alloys [17]. Synthesis of anisotropic precursor powders are key in order to achieve high remanence as well as high $(BH)_{max}$. This was obtained through complicated synthesis routes, such as field aligned resin binding [21], hot pressing [22] and spark plasma sintering [23] etc., The magnetic properties of MnBi are largely governed by the fractional presence of various phases, compositions and microstructures. Hence, detailed structural and compositional studies are essential to improve the magnetic properties of MnBi [24].

In this study, we have reported a simple route to synthesize potential permanent magnetic material with reasonable saturation moment and magnetic energy product. The LTP-FM MnBi sample is synthesized by solid state reaction route, heating at 310°C for 48 hours followed by rapid quenching. *XRD* patterns reveal the formation of MnBi as main phase with some minor Bi and γ-Mn impurities. The magnetic properties are studied by M-H characterization, which showed a magnetic moment of ~62emu/g and $(BH)_{max}$ of 4.01MGOe at room temperature.

## 2. Experimental

The bulk MnBi is synthesized through vacuum encapsulation solid state reaction route. The initial atomic composition Mn: Bi was taken in 1:1 ratio and ground thoroughly. After initial grinding, the samples are pressed into rectangular pellet form and sintered at various temperatures and for variable durations. The optimum moment and phase purity is obtained for the sample sintered at 310°C in vacuum for 48h and rapid quenched. Phase formation and purity of the samples are characterized by using Rigaku X-ray diffractometer (Cu-K$_α$ line) at room temperature. The phase purity and lattice parameter refining are carried out by Rietveld refinement programme (Fullprof version). Micro-structural characterization of the sample was carried out using (SEM) *ZEISS* EVO MA-10 Scanning Electron Microscope. The magnetization measurement is carried out on Magnetic Property Measurement System (MPMS-7T).

## 3. Results and discussion

Fig. 1(a) depicts the *XRD* pattern of the MnBi samples processed at various temperatures. It is observed that the samples heat treated at 310$^o$C for 48h exhibits the least elemental impurity. The impurities observed within main phase MnBi are elemental Bi and Mn. The peaks corresponding to Bi are marked in XRD pattern. The peaks associated with Mn are difficult to be assigned on XRD pattern because most of them overlap with major phase MnBi peaks. The magnetic transition from LTP to HTP takes place at the Curie temperature of 360$^o$C. So, we quenched the sample at 310$^o$C to obtain LTP MnBi [10, 16]. It is clear from the XRD data (Fig. 1a) that the samples heat treated at slightly higher or lower temperatures contain larger amounts of impurities. This suggests that heating schedule has a lot of impact on the phase purity of the synthesized LTP-FM MnBi phase. Also, it was observed that Bi comes out from the samples, which are sintered above the peritectic temperature (>446$^o$C). Rietveld refinement is performed to simulate the *XRD* pattern of the least impurity containing samples, i.e., the ones being heat treated at 310$^o$C for 48h and 72h [Fig. 1(b)]. The MnBi majority phase is crystallized in hexagonal *P63/mmc* space group (NiAs type structure). The refined lattice parameters are found to be $a$ =4.288 (2) Å and $c$ =6.118 (4) Å for 48h sintered sample. The Mn atom occupies *2a* site (0, 0, 0), whereas the Bi is positioned at *2c* site (1/3, 2/3, 1/4). The representative unit cell is shown in Fig. 1(c).

Fig.2 (a) and (b) show the *SEM* images of MnBi samples sintered at 310$^o$C and 500$^o$C respectively. Figure 2(a) shows the formation of granular majority MnBi phase with average grain size of around 1μm. On the other hand the microstructure in Figure 2(b) consists of minor MnBi phase with major unreacted Bi and Mn phases, with plate like structures [20]. It is observed that the surface morphology of sample sintered at 310$^o$C is more homogeneous, with nearly uniform size distribution of the MnBi particles than the one being sintered at 500$^o$C. This is in accordance with *XRD* data, which shows better phase formation for 310$^o$C sintered samples. It is clearly evident that the grain size is more uniform with better MnBi phase for 310$^o$C sintered samples (Fig. 2a), whereas for 500$^o$C sample the un-reacted Bi and Mn are observed along with MnBi (Fig. 2b). The elemental composition of MnBi is analytically examined from the area scanned energy dispersive X-ray (*EDX*) measurements.

Fig.2(c) shows the atomic composition of LTP-FM MnBi sintered at 310$^o$C, confirming approximately the intended atomic ratios as nominal MnBi composition.

Fig. 3(a) exhibits the room temperature (300K) *M-H* plots for the samples sintered at various temperatures i.e., 300$^o$C/48h, 310$^o$C/48h and 310$^o$C/72 h. The inset shows the extended *M-H* loop of the same in applied field of up to 6Tesla. The samples sintered at higher temperatures i.e. above the peritectic temperature, having higher impurity showed lower coercivity and saturation moment (not shown in fig). It can be seen from Fig. 3(a) that the sample synthesized at 310$^o$C/48h has optimum performance in terms of coercivity and saturation moment. Fig 3(b) shows the zoomed *M-H* plot of the sample sintered at 310$^o$C /48h, where the intersection of permeance line with the hysteresis loop is drawn and the magnetic energy product ($BH_{max}$) is calculated to be 4.01MGOe. The theoretically predicted value for LTP-FM MnBi is around 18MGOe [23]. Certainly our experimentally observed value is less than one fourth to the theoretical value. However, our results are quite comparable with the experimentally observed reported value of around 4.3 MGOe [22] for MnBi being processed employing different routes. It is notable that the studied samples are synthesized by single step vacuum encapsulation technique in comparison to the reported specialised techniques of references [3, 22, 23]. Further the *M-H* loops of presently optimized MnBi exhibits some misalignment of Mn moments in matrix. This leaves further scope for completely aligning the Mn moments in matrix and thus achieving even higher values of saturation moment and the magnetic energy product.

In conclusion, we have optimized (vacuum seal and 310$^o$C/48h quenched) synthesis conditions for MnBi compound. The structural refinement showed hexagonal structural crystallization of MnBi as major phase along with some minor impurities. The compound exhibits appreciable magnetic moment (40emu/gram, 1Tesla) and energy density product (4.01MGOe) at room temperature.

**Acknowledgements**

This work was carried out under CSIR (India) network project PSC-0109. The authors would like to thank DNPL Prof. R. C. Budhani for his constant support and encouragement. Authors are also thankful to Mr. A. K. Sood and Mr. Jai Tawale for SEM measurements. Dr. Anurag Gupta is acknowledged for carrying out the magnetic measurements. One of the authors S. K. S. would like to acknowledge CSIR, India for providing the fellowship.

**Figure Caption**

**Fig. 1 (a):** XRD pattern of MnBi samples with variation of sintering temperature and time. Impurity phases are marked with *.

**Fig. 1 (b):** Rietveld fitted XRD pattern of MnBi samples with space group *P63/mmc* sintered at $310^{o}$C /48h, $310^{o}$C /72 h. Impurity phases are marked with *.

**Fig. 1 (c):** Représentative unit cell of MnBi in *P63/mmc* crystallization

**Fig. 2 (a):** SEM images of MnBi sintered at $310^{o}$C/48h; **(b)** MnBi sintered at $500^{o}$C/48h and **(c)** Area scanned EDX of the sample sintered at $310^{o}$C/48h.

**Fig. 3(a):** Room temperature (300K) *M-H* plots measurement of MnBi samples sintered at $300^{o}$C/48h, $310^{o}$C/48h, $310^{o}$C/72 h. The inset of the figure shows the same in extended field region up to 6 T.

**Fig. 3(b):** Room temperature (300K) *M-H* plots measurement of MnBi samples sintered at $310^{o}$C /48h showing the calculated $BH_{max}$ plot.

Fig. 1 (a)

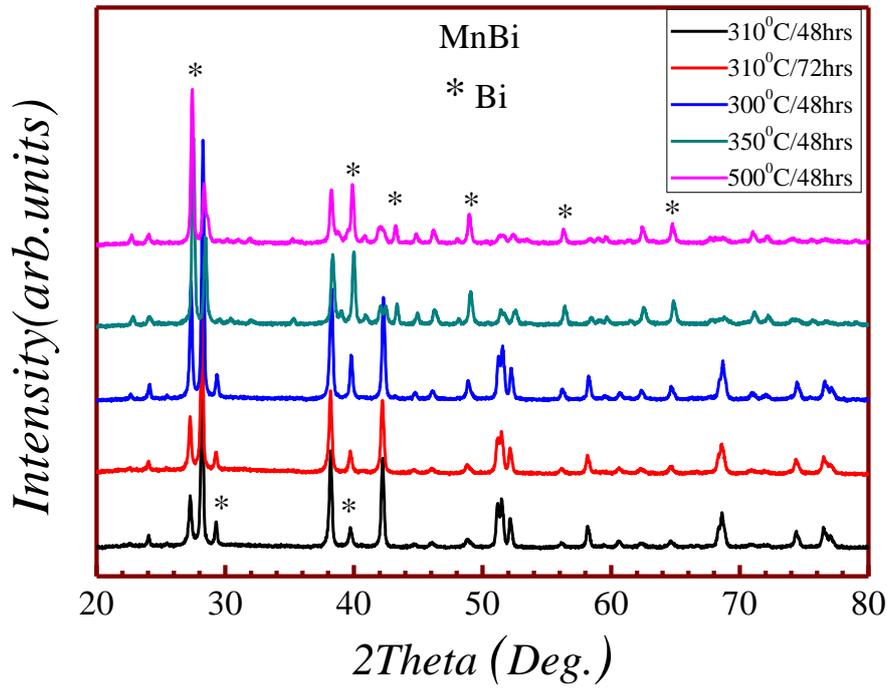

Fig.1 (b)

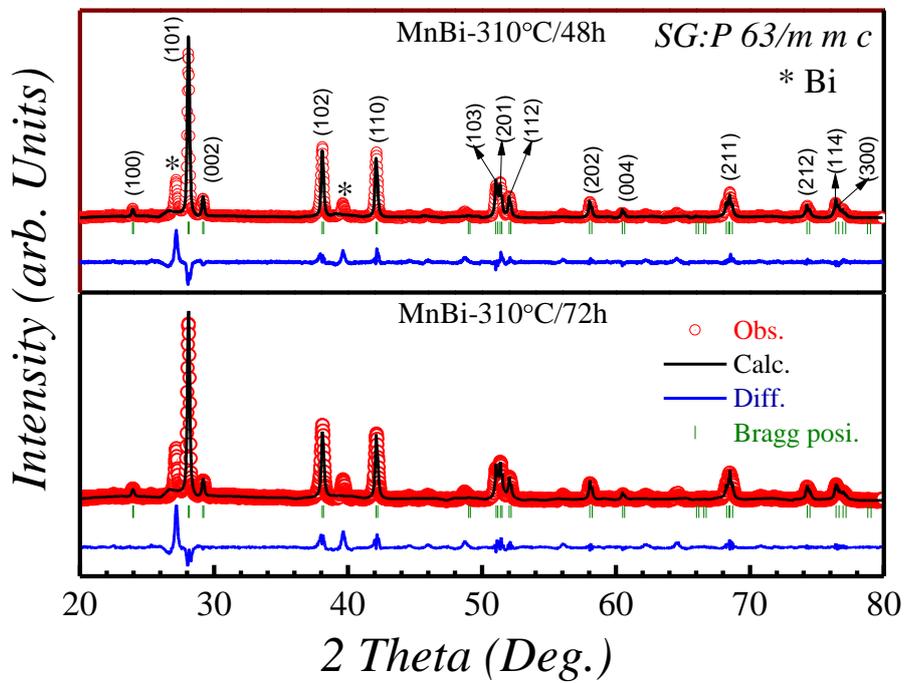

Fig.1 (c)

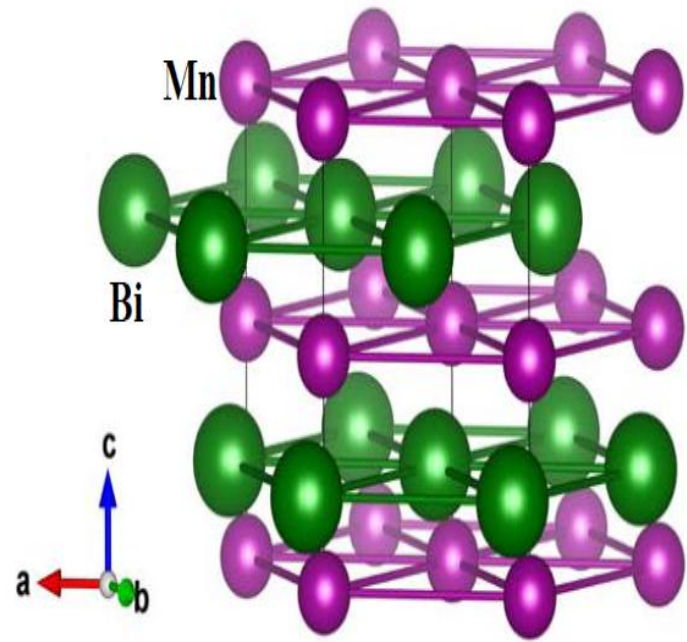

Fig. 2(a) and (b)

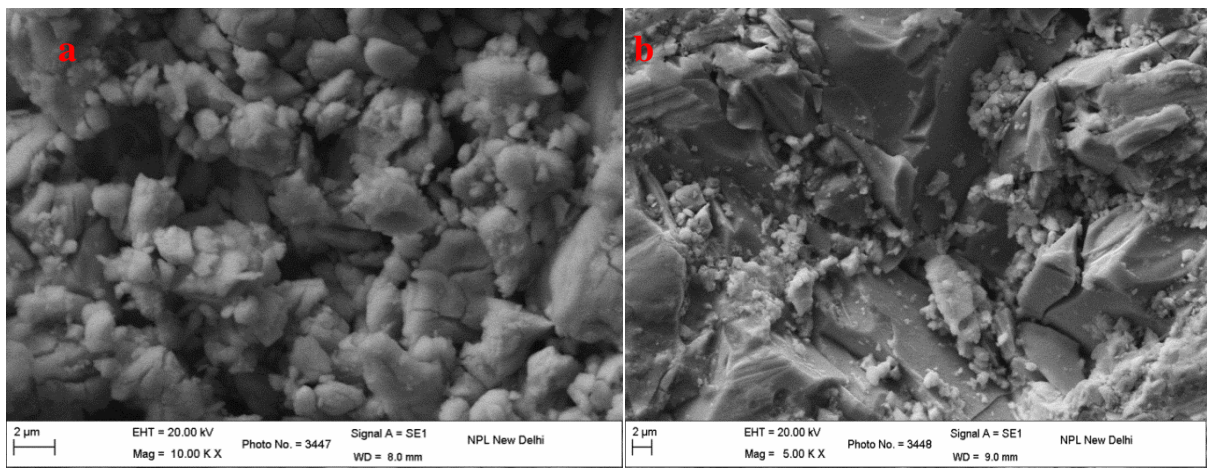

Fig. 2 (c)

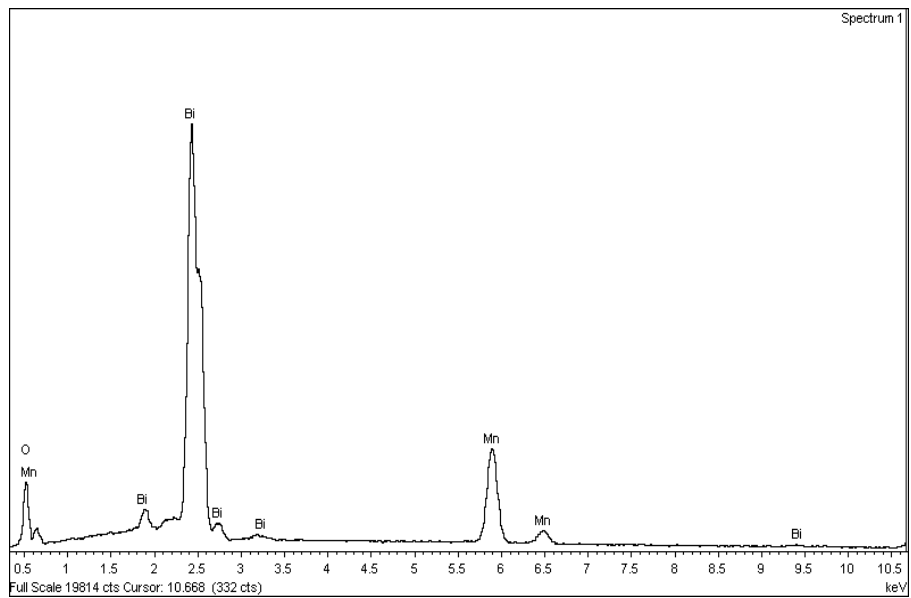

Fig. 3(a)

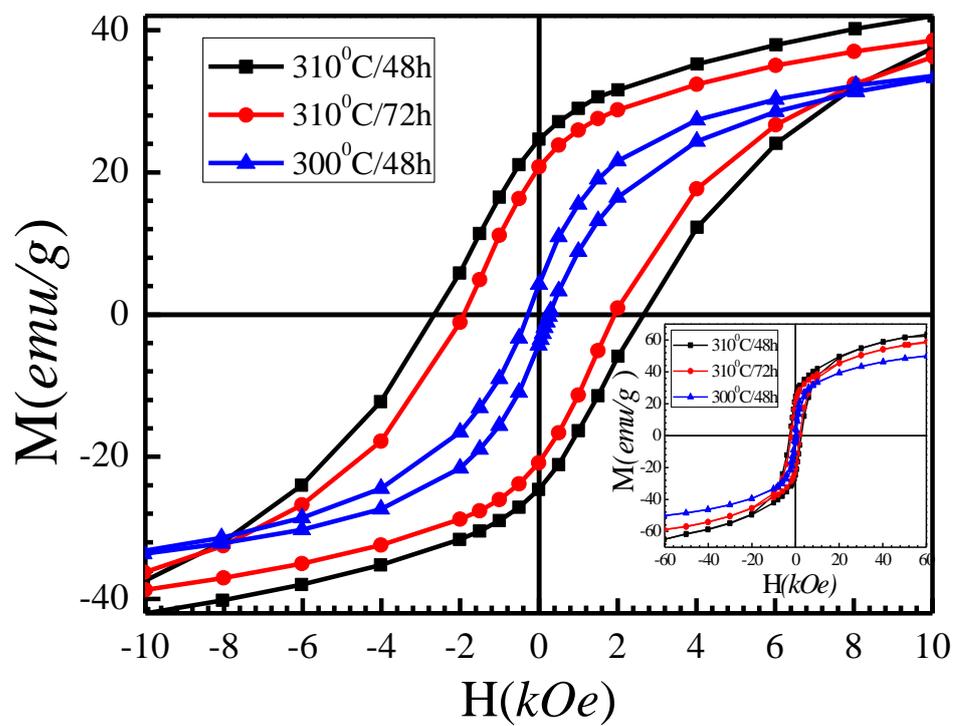

Fig 3 (b)

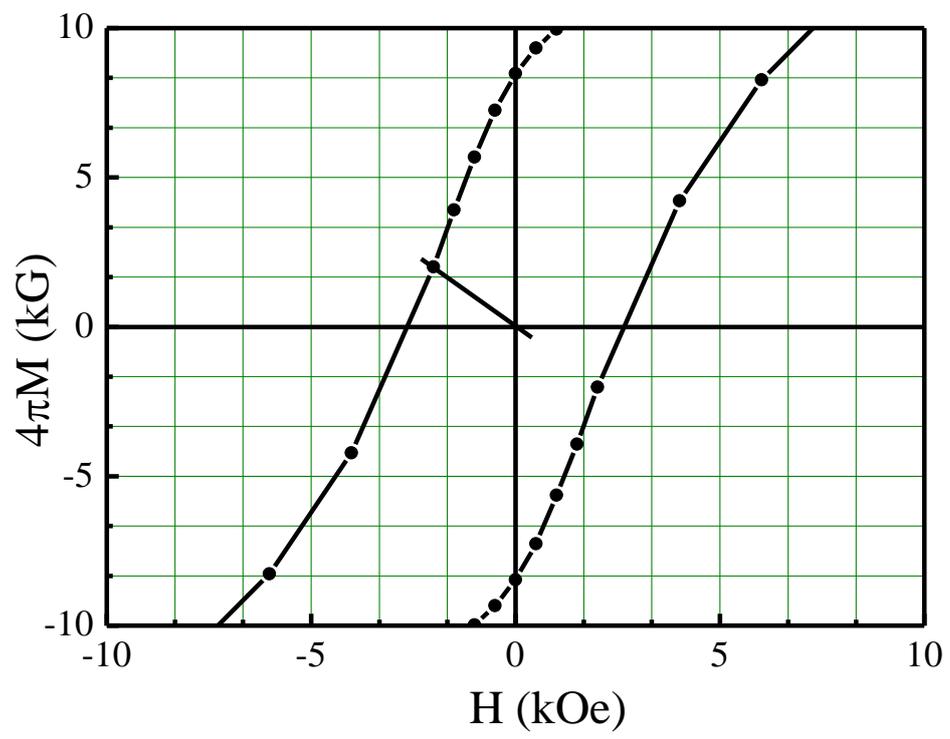